\documentclass[aip,jcp]{revtex4-1}

\usepackage{amsmath}
\usepackage{graphicx}

\newcommand{\qvec}{{\bf q}}
\newcommand{\pvec}{{\bf p}}
\newcommand{\fvec}{{\bf f}}

\begin{document}

\title{Stabilization of {\it Ab Initio} Molecular Dynamics Simulations at Large Time Steps}

\author{Eiji  Tsuchida \\ 
   Nanosystem Research Institute, \\
   National Institute of Advanced Industrial Science and Technology (AIST), \\
   Tsukuba Central 2, Umezono 1-1-1, Tsukuba 305-8568, Japan }





\begin{abstract}
The Verlet method is still widely used to integrate the equations of motion 
in {\it ab initio} molecular dynamics simulations. 
We show that the stability limit of the Verlet method 
may be significantly increased by setting an upper limit on the kinetic energy of each atom 
with only a small loss in accuracy. The validity of this approach is demonstrated 
for molten lithium fluoride. 
\end{abstract}


\maketitle

\section{Introduction}
Researchers in many fields of science and technology now routinely use 
{\it ab initio} molecular dynamics (AIMD) simulations for investigating various properties of 
complex systems \cite{AIMD1}. 
However, the computational cost of AIMD is still a serious obstacle, 
even on a supercomputer. If, however, the purpose of the simulation is to obtain  
low-energy conformations through simulated annealing, or to equilibrate the system 
prior to the production run, the accuracy of time integration is not of primary concern. 
In this case, the computational cost of AIMD is minimized by using the largest possible 
time step. When the Verlet method is used to integrate the equations of motion, 
the maximum size of the time step is given by $h_{\rm max}=T/\pi$, 
where $T$ is the period of the fastest oscillation in the system \cite{CLREV1}. 
In practice, however, AIMD simulations often break down at $h \sim h_{\rm max}/2$ 
because of the strong anharmonicity of the interatomic forces. 
In this work, we show that a slight modification of the Verlet method 
allows us to increase the stability limit of the time step significantly with only a small loss in accuracy. 

\section{Theory}
\label{THEOSEC}

The classical Hamiltonian for a system of $N$ atoms is given by 
\begin{equation}
\label{OHAM}
H (\qvec,\pvec) = \frac12 \, \pvec^T M^{-1} \pvec + U(\qvec), 
\end{equation}
where $\qvec$ and $\pvec$ are vectors of dimension $3N$, 
representing atomic positions and momenta, 
$M$ is the mass matrix, and $U(\qvec)$ is the potential energy. 
Then, $\qvec (t)$ and  $\pvec (t)$ satisfy the equations of motion, 
\begin{eqnarray}
 & & \frac{d \qvec}{dt} =  \frac{\partial H}{\partial \pvec} = M^{-1} \pvec \\
 & & \frac{d \pvec}{dt} = -\frac{\partial H}{\partial \qvec} = -\frac{\partial U}{\partial \qvec}. 
\end{eqnarray}
In general, these equations cannot be solved analytically, and thus must be evaluated numerically. 
When these equations are discretized in time with a time step of $h$, 
and neglecting $O(h^3)$ terms, the (velocity) Verlet method is obtained \cite{STATXT}: 
\begin{eqnarray}
& &\pvec_{n+\frac12} = \pvec_n + \frac{h}{2} \, \fvec_n \\
& &\qvec_{n+1}       = \qvec_n + h M^{-1} \pvec_{n+\frac12}\\
& &\pvec_{n+1}       = \pvec_{n+\frac12} + \frac{h}{2} \, \fvec_{n+1},
\end{eqnarray}
where the force is defined by $\fvec = - \partial U / \partial \qvec$, and 
the subscript denotes the time-step number. 
This integrator is symplectic, time-reversible, and 
requires only one force evaluation per step. 
Therefore, the Verlet method is still widely used for AIMD \cite{FNOTE1}. 
It is common practice to use $h < h_{\rm max}/5$ for production runs, 
where $h_{\rm max}$ is the theoretical limit defined in the Introduction. 
In contrast, much larger time steps are acceptable 
for equilibration and simulated annealing where only modest accuracy is required. 
At some point, however, the total energy diverges and time evolution breaks down. 
In our experience, the breakdown occurs at $h \sim h_{\rm max}/2$ in the following manner. 
\begin{description}
\item[(a)] Two atoms approach each other very closely.
\item[(b)] Strong repulsive forces act between them. 
	This effect is more pronounced in AIMD because of the stronger anharmonicity. 
\item[(c)] These forces give rise to large atomic velocities. 
\item[(d)] Go to (a) if necessary. 
\end{description}
When the time step is large, this cycle often continues 
until two atoms nearly overlap, indicating the breakdown of the simulations. 
We also note that even a single atom can cause a breakdown if its kinetic energy is sufficiently large. 

The basic idea of our approach is to avoid the breakdown 
by setting an upper limit on the kinetic energy of each atom. 
To this end, we propose to modify the Verlet method as follows: 
\begin{eqnarray}
& &\pvec_{n+\frac12} = \pvec_n + \frac{h}{2} \, \fvec_n \\
\label{MODPEQ}
& &\mbox{Modify ($\pvec_{n+\frac12}$)} \\
& &\qvec_{n+1}       = \qvec_n + h M^{-1} \pvec_{n+\frac12}\\
& &\pvec_{n+1}       = \pvec_{n+\frac12} + \frac{h}{2} \, \fvec_{n+1},
\end{eqnarray}
where the modification of $\pvec$ at $t=(n+\frac12)h$, Eq.(\ref{MODPEQ}), can be written as
\begin{eqnarray}
& & \mbox{for}\,\,i=1,N \\
\label{EKINCOND}
& & \quad \quad \mbox{if}\,\,(E_{\rm kin} (i) > E_{\rm cut})\,\,\mbox{then} \\
& & \quad \quad \quad \quad \pvec (i) = \pvec (i) \times \gamma_i \\ 
& & \quad \quad \mbox{endif} \\
& & \mbox{end do}
\end{eqnarray}
in pseudo-code format. Here, $\gamma_i$ is defined by 
\begin{equation}
\gamma_i = \sqrt{\frac{E_{\rm cut}}{E_{\rm kin} (i)}} \times \beta
\end{equation}
with 
\begin{eqnarray}
& & E_{\rm kin}(i)=\frac{\pvec (i)^2}{2 m_i} \\
& & E_{\rm cut}=\frac{3}{2} k_{\rm B} T_0 \times \alpha^2, 
\end{eqnarray}
and $T_0$ is the target temperature. 
This procedure requires two dimensionless parameters: 
$\alpha$ determines the cutoff energy and $\beta$ corresponds to the kinetic energy after the scaling, i.e. 
\begin{equation}
E_{\rm kin}^{\rm new} (i) = E_{\rm cut} \times \beta^2 = \mbox{const.}
\end{equation}
holds for all atoms which satisfy Eq.(\ref{EKINCOND}). 
In what follows, this procedure is called {\it stabilization}. 
It is also possible to apply the stabilization to thermostatted systems without serious difficulties. 
Moreover, the computational cost is negligible. 

On the other hand, the current implementation ignores the conservation of the total energy and momentum. 
When a thermostat is applied, this is not a serious problem 
as long as only a small fraction of the atoms satisfy Eq.(\ref{EKINCOND}) at each time step. 
If, however, the drift of the total energy is significant, 
it may be necessary to include dissipative forces to compensate for the drift \cite{NF1,NF2,NF3,NF4}.

\section{Results}

Here we study the effect of stabilization on the performance of AIMD simulations 
for a high-temperature molten salt. 
Molten lithium fluoride was modeled by 72 LiF pairs in a cubic supercell of length 12.06 \AA. 
Atomic forces were calculated within the density functional theory \cite{HK,KS,PBE}, and 
norm-conserving pseudopotentials were used \cite{GTH,HGH}. 
The electronic orbitals were expanded by the 
finite-element basis functions \cite{FEM1,FEM2} with 
an average cutoff energy of 78 Ryd, while the resolution was enhanced by more than a factor of two 
near the atoms \cite{ACC}. 
Only the $\Gamma$-point was used to sample the Brillouin zone. 
The electronic states were quenched to the ground state at each time step 
with the limited-memory BFGS method \cite{LINO,QNFEM,MIXP}. 
The equations of motion were integrated using the Verlet method with and without the stabilization. 
After equilibration, production runs of 240 ps were carried out using $h =4-11$ fs. 
The temperature was controlled by the Berendsen thermostat 
with a relaxation time of $\tau$. 
In Table \ref{MDLIF}, we show the simulation details for all runs. 
We used the same initial conditions $(\qvec_0, \pvec_0)$ and experimental masses for all atoms in these runs. 
We note in passing that the period of the fastest oscillation in this system is not a well-defined quantity. 
However, $h =$ 0.5 fs \cite{SRSSBM}, 1.5 fs \cite{BRMR}, and 4 fs \cite{CJM} were used 
in previous studies of this system. 

\begin{figure}
  \begin{center}
  \includegraphics[width=9cm]{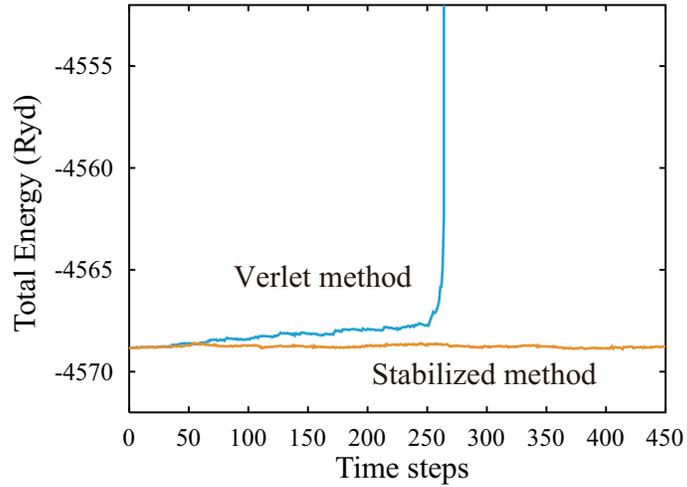}
  \end{center}
  \caption{Time evolution of the total energy for $h =$ 8 fs with and without the stabilization.}
  \label{STABFIG}
\end{figure}

\begin{figure}
  \begin{center}
  \includegraphics[width=15cm]{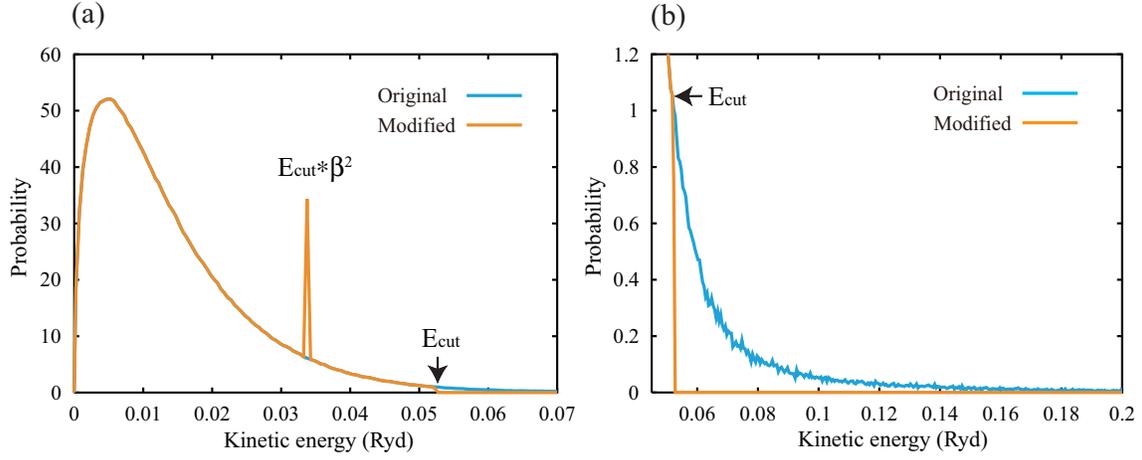}
  \end{center}
  \caption{Distributions of the kinetic energy of each atom at $t=(n+\frac12)h$ for $h =$ 10 fs.}
  \label{EKINFIG}
\end{figure}

\begin{figure}
  \begin{center}
  \includegraphics[width=15cm]{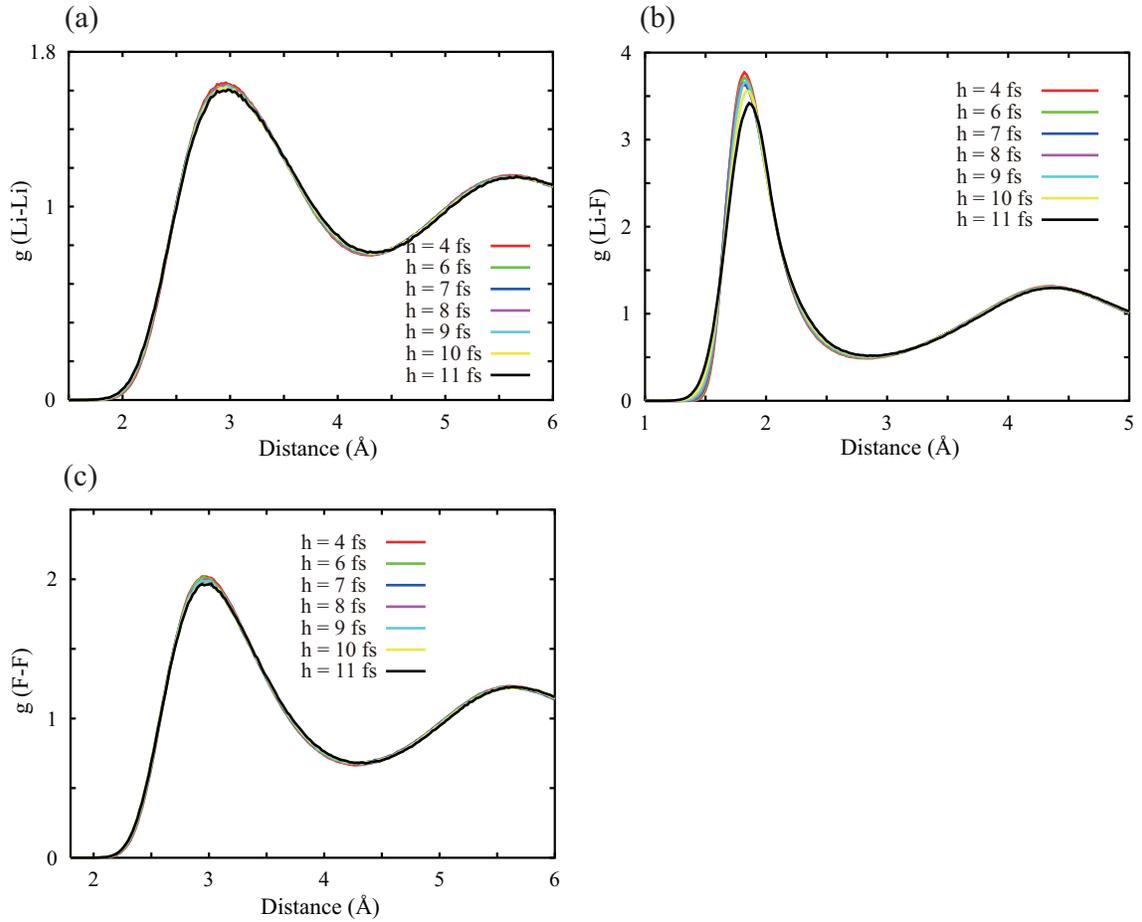}
  \end{center}
  \caption{Radial distribution functions for molten lithium fluoride: 
  (a) $g($Li-Li$)$, (b) $g($Li-F$)$, and (c) $g($F-F$)$. }
  \label{RDFFIG}
\end{figure}

The Verlet method was found to be stable up to $h =$ 6 fs without stabilization, 
while a divergence of the total energy was observed at $h =$ 7 fs after running for 203 ps. 
When the stabilization was performed, the simulation was valid even for $h =$ 11 fs. 
We note, however, that the values of $\alpha$, $\beta$, and $\tau$ had to be reduced 
for larger $h$ to stabilize the simulations. 
We show the effect of stabilization for $h =$ 8 fs in Fig.\ref{STABFIG}. 
Distributions of the kinetic energy before and after 
the stabilization are compared in Fig.\ref{EKINFIG}. 
The original distribution decays very slowly with energy, and is extended up to 4.4 Ryd. 
This {\it long tail} is responsible for the breakdown of the simulations. 
After the stabilization, the distribution is truncated at $E_{\rm cut}$. 
In Fig.\ref{RDFFIG}, we compare the radial distribution functions 
($g($Li-Li$)$, $g($Li-F$)$, and $g($F-F$)$) obtained from the simulations. 
The first peak of $g($Li-F$)$ shows some broadening for $h =$ 10 and 11 fs. 
However, all runs give similar results at larger distances. 
Moreover, $g($Li-Li$)$ and $g($F-F$)$ RDFs remain essentially the same for all runs up to $h =$ 11 fs. 
The self-diffusion coefficients given in Table \ref{MDLIF} show some scatter, 
but no clear dependence on the simulation conditions \cite{FNOTE2}. 
These results are in reasonable agreement with the experimental values 
(8.9$\times$10$^{-5}$cm$^2$/s for Li and 7.2$\times$10$^{-5}$cm$^2$/s for F) 
measured at 1123 K \cite{SRSSBM}. 

\clearpage

\section{Conclusion}
We have shown that the stability limit of the Verlet method can be increased by $\sim 50 \%$ 
for molten LiF without significant loss in accuracy 
if the kinetic energy of each atom is carefully controlled. 
Preliminary AIMD simulations of liquid water are also showing promising results. 
The stabilization method presented in this paper would be particularly useful 
when only modest accuracy is required 
within the framework of AIMD, e.g., for equilibration and global optimization. 
This algorithm may also be used in conjunction with other methods to 
accelerate the simulations even further, such as the Langevin dynamics \cite{NF1,NF2,NF3,NF4}, 
linear scaling method \cite{AOMM,OZ,ONREV} 
and mass scaling method \cite{MTMD2}.

\section*{Acknowledgments}
This work has been supported by the Strategic Programs for Innovative Research (SPIRE) 
and a KAKENHI grant (22104001) from the Ministry of Education, Culture, Sports, 
Science \& Technology (MEXT), 
and the Computational Materials Science Initiative (CMSI), Japan.

\clearpage

\begin{table}
\caption{Details of AIMD simulations for molten LiF. 
Simulation lengths shorter than 240 ps indicate failed runs. 
MOD represents the probability that each atom satisfies Eq.(\ref{EKINCOND}) 
at each time step. 
}
\label{MDLIF}
\begin{center}
\begin{tabular}{cccccccccc}
\hline
\hline
\ \ $h$ \ \ & Length & Stabilization & \ \ \ $\tau$ \ \ \  & \ \ \ $\alpha$ \ \ \ & \ \ \ $\beta$ \ \ \ & 
MOD & $D_{\rm self}$(Li) & $D_{\rm self}$(F) & Temperature \\
(fs) & (ps) & & (ps) & & & (\%) & (10$^{-5}$ cm$^2$/s) & (10$^{-5}$ cm$^2$/s) & (K) \\
\hline
4  & 240 & No  & 0.4   & - & - & - & 11.4 & 7.2 & 1248.3\\
6  & 240 & No  & 0.4   & - & - & - & 13.8 & 9.7 & 1256.0\\
7  & (203) & No  & 0.4   & - & - & - & 15.9 & 9.1 & 1298.9\\
8  &   (2) & No  & 0.4   & - & - & - & - & - & -\\
8  & 240 & Yes & 0.4   & 2.3 & 0.9 & 0.2 & 12.6 & 8.5 & 1255.6\\
9  & 240 & Yes & 0.133 & 2.2 & 0.8 & 0.5 & 11.8 & 8.3 & 1246.3\\
10 & 240 & Yes & 0.133 & 2.1 & 0.8 & 1.4 & 12.9 & 7.7 & 1270.2\\
11 & 240 & Yes & 0.133 & 2.0 & 0.8 & 3.2 & 11.0 & 7.0 & 1279.7\\
\hline
\hline
\end{tabular}
\end{center}
\end{table}

\begin{thebibliography}{9}
\bibitem{FNOTE1} In this work, we focus on the Born-Oppenheimer dynamics using tight convergence criteria.
\bibitem{FNOTE2} When the stabilization is performed according to Sec.\ref{THEOSEC}, 
	the conservation of total momentum is not strictly valid. 
	Therefore, the motion of the center of mass was explicitly taken into account 
	when calculating the self-diffusion coefficients.
\end{thebibliography}
\end{document}